\documentclass[prl,twocolumn,showpacs,amsmath,amssymb,superscriptaddress,groupedaddress]{revtex4-1}

\usepackage{graphicx}
\usepackage{ifthen}
\usepackage{hyperref}
\hypersetup{
 pdfnewwindow=true, colorlinks=true,
 linkcolor=blue, anchorcolor=blue,
 citecolor=blue, filecolor=blue,
 menucolor=blue, urlcolor=blue}
\usepackage{breakurl}
\usepackage{dcolumn}
\newcolumntype{.}{D{.}{.}{1}}
\newcommand{\mc}[1]{\multicolumn{1}{c@{}}{#1}}

\usepackage{pdfpages}
\usepackage{pgffor}

\begin{document}

\title{Large electron-phonon interactions from FeSe phonons in a monolayer}

\author{Sinisa Coh}
\email{sinisacoh@gmail.com}
\author{Marvin L. Cohen}
\author{Steven G. Louie}
\affiliation{Department of Physics, University of California at
  Berkeley and Materials Sciences Division, Lawrence Berkeley National
  Laboratory, Berkeley, California 94720, USA}

\date{\today}

\begin{abstract} 
  We show that electron-phonon coupling can induce strong electron
  pairing in an FeSe monolayer on a SrTiO$_3$ substrate (experimental
  indications for superconducting $T_{\rm c}$ are between 65 and
  109~K). The role of the SrTiO$_3$ substrate in increasing the
  coupling is two-fold.  First, the interaction of the FeSe and
  TiO$_2$ terminated face of SrTiO$_3$ prevents the FeSe monolayer
  from undergoing a shear-type (orthorhombic, nematic) structural
  phase transition.  Second, the substrate allows an
  anti-ferromagnetic ground state of FeSe which opens electron-phonon
  coupling channels within the monolayer that are prevented by
  symmetry in the non-magnetic phase.  The spectral function for the
  electron-phonon coupling ($\alpha^2F$) in our calculations agrees
  well with inelastic tunneling data.
\end{abstract}


\maketitle

Small variations of external perturbations can result in the favoring
of one of a range of competing structural, electronic, and magnetic
ground states for FeSe. In particular, the superconducting transition
temperature in FeSe is reputed to vary from almost 0~K when slightly
Fe doped~\cite{mcqueen2009} to 65~K when placed in a monolayer form on
a SrTiO$_3$ substrate~\cite{wang2012,jjlee2013,liu2012,shaolong2013},
and transport measurements from a recent work~\cite{ge2014} indicate
an even larger $T_{\rm c}$, close to 109~K.  Although FeSe has a
simpler structure to the other iron-based superconductors it resembles
components of their structure, and there is the possibility that the
mechanism responsible for high temperature superconductivity in
monolayer FeSe may extend to other iron-based compounds.

Early calculations (Refs.~\cite{boeri2008,mazin2008}) based on density
functional theory (DFT) estimated electron-phonon coupling in the
iron-based superconductors to be at least 5-6 times too small to
explain the transition temperatures found experimentally.  Therefore,
a large part of the theoretical and the
experimental~\cite{imai2009,lumsden2010,hanaguri2010} work on
iron-based superconductors in the literature focused on alternative
electron pairing mechanisms such as those associated with magnetic
fluctuations. In this letter we suggest that the early first-principle
calculations may have underestimated the electron-phonon coupling in
FeSe, and we conclude that conventional electron-phonon coupling may
be strong enough to contribute significantly to the
electron pairing in an FeSe monolayer on SrTiO$_3$ and perhaps other
iron-based superconductors.

We focus here on an FeSe monolayer on a TiO$_2$ terminated SrTiO$_3$
substrate. We show that the interaction between the substrate and the
FeSe monolayer leads to a high phonon-mediated superconducting T$_{\rm
  c}$ by providing a {\it structural template} which holds FeSe near
its structural and magnetic phase transitions. When this structural
template is not present (as in bulk FeSe or a monolayer of FeSe on a
weakly interacting substrate) the system condenses to a different
ground state (orthorhombic and non-magnetic) with a reduced
electron-phonon coupling.

Among the many possible ground states of FeSe, calculations
  based on a semi-local density approximation (GGA) to the DFT
select a ground state inconsistent with structural~\cite{subedi2008},
electronic~\cite{nakayama2010,tamai2010,chen2010,maletz2014}, and
magnetic~\cite{li2009,subedi2008} measurements.  While the
shortcomings of standard GGA bands for transition metals (such as Fe)
can often be corrected by semi-empirically including a Hubbard or a
Hund interaction (as in the GGA+U method,~\cite{liechtenstein1995}),
this is not the case for FeSe.\cite{fawei2013} Higher
levels of theory, such as GW or DMFT in
Refs.~\cite{yin2011,tomczak2012}, can correctly reproduce most
electronic properties of bulk FeSe; however, calculation of the
electron-phonon coupling with these methods relies on a simplified
deformation potential approximation, as in Ref.~\cite{mangal2014}
since electron-phonon coupling matrix elements are difficult to
obtain.

\begin{figure}[!t]
\centering
\includegraphics{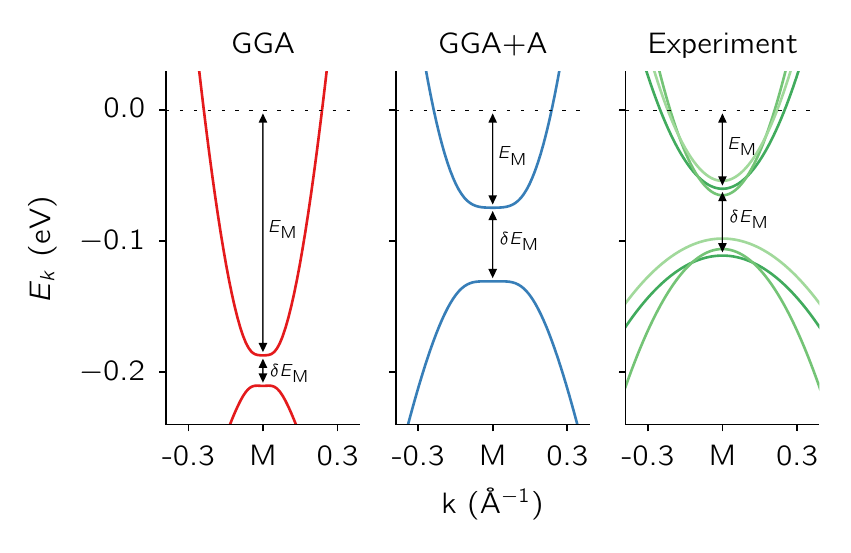}
\caption{\label{fig:bs} Electronic band structure near the M point
  (full band structure is shown in the
    supplement~\cite{supplement}) of monolayer FeSe on SrTiO$_3$ in
  GGA (red), GGA+A with $A=A_{\rm c}$ (blue), and experimental
  results. We fit the different experimental data to a parabola
  (light~\cite{jjlee2013}, medium~\cite{liu2012}, and dark
  green~\cite{shaolong2013}).}
\end{figure}

\begin{table}[!t]
  \caption{\label{tab:comparison} A comparison of the magnetic moment
    on the iron atom ($\mu$), shear angle $\alpha$ (measured between
    the primitive unit cell vectors $a$ and $b$), top of the $\Gamma$
    band ($E_{\Gamma}$) and bottom of the M band (occupied bandwidth,
    $E_{\rm M}$) relative to the Fermi level, and the band splitting
    at the M point ($\delta E_{ \rm M}$) in GGA, GGA+A using $A=A_{\rm
      c}$, and from experiments
    (Refs.~\cite{mcqueen2009,li2009,jjlee2013,liu2012,shaolong2013}).
    Parameter $A$ is tuned to $A=A_{\rm c}$ so that occupied bandwidth
    of the M-point electron pocket ($E_{\rm M}$) agrees with
    experimental data.  However, using $A=A_{\rm c}$ significantly
    improves other properties of FeSe as well.}
 \begin{ruledtabular}
 \begin{tabular}{lc.ccccc}
 & \multicolumn{2}{c}{Bulk} & &\multicolumn{4}{c}{Monolayer on SrTiO$_3$} \\
  \cline{2-3} \cline{5-8}
& \mc{$\mu$\footnotemark[1]} & \mc{$\alpha$\footnotemark[2]} &
& \mc{$\mu$} & \mc{$E_{\Gamma}$} & \mc{$E_{\rm M}$} & \mc{$\delta E_{ \rm M }$}
\\
& \mc{($\mu_{\rm{B}}$)} & \mc{($^{\circ}$)} &
& \mc{($\mu_{\rm{B}}$)} & \mc{(eV)} & \mc{(eV)} & \mc{(eV)}
\\
\hline
GGA        & 2.4 & 90    & &  2.6 &  0.66 &  0.19 & 0.02 \\
GGA+A ($A_{\rm c}$)
           & 0   & 89.96 & &  1.8 &  0.17 &  0.07 & 0.06 \\
Experiment & 0   & 89.7  & &  \footnotemark[3]
                                  &  0.08 &  0.06 & 0.05 \\
    \end{tabular}
\footnotetext[1]{Using experimental crystal structure.}
\footnotetext[2]{Fully relaxed with the van der Waals correction
     from Ref.~\cite{grimme2006}.}
\footnotetext[3]{Unknown.}
  \end{ruledtabular}
\end{table}

Here we show that making the potential on the iron atoms slightly more
repulsive for electrons renormalizes the bands near the Fermi level
and selects a ground state of FeSe consistent with most experimental
data. More specifically, in this method (GGA+A), we
empirically\footnote{This approach is similar in spirit to the
  empirical pseudopotential method from Ref.~\cite{cohen1970} and the
  semi-empirical method from Ref.~\cite{wang1995}.}  replace the
potential $V_{\rm GGA} ({\bf r})$ within the semi-local density
approximation (GGA) with
\begin{align}
  V_{\rm GGA} ({\bf r}) + A \sum_i f(|{\bf r}-{\bf r}_i|).
  \label{eq:Af}
\end{align}
The idea here is to mitigate empirically the fact that the GGA
exchange-correlation potential is not the self energy without the second
term in Eq.~\ref{eq:Af}.  We find that the detailed form of the
dimensionless function $f(r)>0$ is irrelevant for the computed
physical properties of FeSe, as long as $f(r)$ is peaked on the Fe
atom (placed at ${\bf r}_i$) and the extent of $f(r)$ is comparable
with the size of the iron atom d-orbital.~\footnote{For example, we
  tried $f(r)\sim r^n \exp(-B r^m)$ with several choices of $0<n<4$,
  $1<m<4$, and $B$ all giving similar results.}  Next, for
a fixed $f(r)$, we tune the parameter $A$ from $0$ up to $A_{\rm c}$ ($>0$)
until~\footnote{We give in the supplement\cite{supplement} the
  numerical value of $A_{\rm c}f(r)$, a list of all parameters used in
  the calculations, and a figure showing the dependence of several
  physical properties of FeSe monolayer on the value of $A$.}  one of
the properties of FeSe (here, occupied bandwidth of the M-point
electron pocket) agrees with
experimental data (compare blue and green curves in
Fig.~\ref{fig:bs}).  Remarkably, using $A=A_{\rm c}$ improves other
salient properties of FeSe as well.  For example, the gap ($\delta E_{
  \rm M}$ in Table~\ref{tab:comparison}) at the bottom of the M
pocket, and the energy of the $\Gamma$ band just below the Fermi level
are improved in the GGA+A, as well as the peak positions in the
density of states at 4 and 6~eV below the Fermi level.~\footnote{In GW
  calculations from Ref.~\cite{tomczak2012} these same peaks near 4
  and 6~eV were found to agree well with the experiment.}  Magnetic
properties are improved as well.  Using the experimental crystal
structure from Ref.~\cite{phelan2009} in both cases, the GGA+A
predicts {\it bulk} FeSe to be nonmagnetic as in experiment, while GGA
predicts large antiferromagnetically aligned magnetic moments $\mu$ on
the iron atoms (favored by 0.5~eV per two Fe atoms over the
non-magnetic ground state). Finally, the crystal structure is improved
in the GGA+A case.  A slight shear present in the experimental
structure as in Ref.~\cite{mcqueen2009} ($\alpha<90^{\circ}$) remains
in the GGA+A approach after the structural relaxation, while it
disappears in the GGA calculations ($\alpha=90^{\circ}$).

In these and subsequent calculations we fixed the doping of FeSe
monolayer to the level of 0.09~electrons per one Fe atom (as found in
ARPES experiments).  In the experiment, this doping likely occurs due
to presence of oxygen vacancies in the SrTiO$_3$ substrate.

Our focus here is on the electron-phonon coupling and
superconductivity in monolayer FeSe. The underlying origin of the
success of the GGA+A is an interesting open question and is left for
future studies. We only note here two points in favor of GGA+A. First,
portion of the electron self-energy $\Sigma({\bf r}, {\bf r}',E)$ that
is missing in the semi-local density approximation is typically large
only when $|{\bf r}-{\bf r}'|$ is comparable to the bond
length,\cite{hybertsen1986} just as for the case of the form of
$f(r)$.  Second, agreement between GGA+A and experiment is improved
not only in monolayer FeSe studied here, but also in bulk KCuF$_3$,
LaNiO$_3$, (La,Sr)$_2$CuO$_4$, SrTiO$_3$ (see
supplement\cite{supplement}), and (Ba,K)Fe$_2$As$_2$.\cite{oh2015}

\begin{figure}[!t]
\centering
\includegraphics{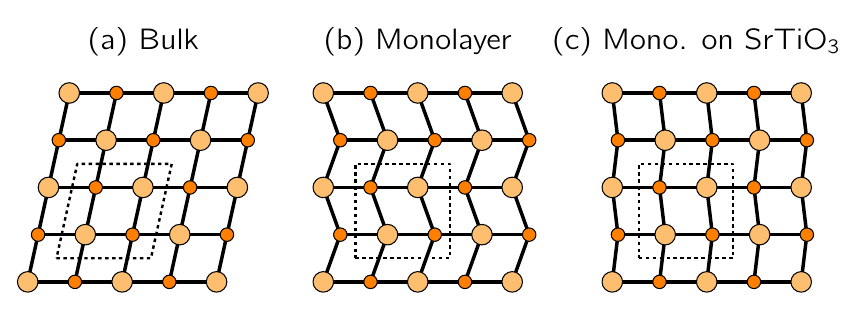}
\caption{\label{fig:dist} Exaggerated structural distortions in FeSe
  bulk, an epitaxially constrained monolayer, and a monolayer on
  SrTiO$_3$.  Small circles are Fe atoms and large circles are Se
  atoms. Primitive unit cell is shown with a dashed gray
    line.}
\end{figure}

Equipped with a better FeSe band structure and ground state than
obtained from a standard GGA calculation, we are now in a position to
compute the electron-phonon coupling strength in the FeSe
monolayer. First we discuss the crystal structure of FeSe used in the
electron-phonon calculation. Bulk FeSe consists of stacked, weakly
interacting, layers of FeSe. Below 90~K these layers are observed to
be slightly sheared as shown in Fig.~\ref{fig:dist}a and discussed in
Ref.~\cite{mcqueen2009} (shear is also present in GGA+A calculation,
but not in GGA). This shear (nematic) distortion is conventionally
described as primitive-tetragonal to base-centered-orthorhombic
structural phase transition.

Since the FeSe layers in bulk are only weakly interacting, we expect
that the tendency towards a shear distortion will be present even in
an {\it isolated} single layer of FeSe. This is indeed what we find in
the case of monolayer FeSe. Even if we epitaxially constrain the
isolated monolayer FeSe unit cell to a cubic SrTiO$_3$ lattice, it
still undergoes a local shear-like structural transition shown in
Fig.~\ref{fig:dist}b (again, only in GGA+A, not in GGA).

However, once FeSe is placed on a TiO$_2$ terminated SrTiO$_3$
substrate, we find that the interaction of Ti and Se atoms together
with the epitaxial strain is able to stabilize FeSe to a nearly square
arrangement (see Fig.~\ref{fig:dist}c and
supplement\cite{supplement}).  A small remnant of the structural
distortion present in FeSe is responsible for the electronic gap
($\delta E_{\rm M}$) at the M point shown in Fig.~\ref{fig:bs} and in
Table~\ref{tab:comparison}.  (An additional smaller component of the
gap results from a built-in electric field between FeSe and SrTiO$_3$,
as discussed in Ref.~\cite{fawei2013}.)  In addition, in the FeSe {\it
  monolayer} on SrTiO$_3$, an antiferromagnetic checkerboard ground
state is preferred by 0.11~eV (per unit cell with two Fe atoms) within
GGA+A over the non-magnetic one, despite the fact that the opposite is
the case for {\it bulk} FeSe.

The main effect of the SrTiO$_3$ on the FeSe is the structural
stabilization described above of a non-sheared and antiferromagnetic ground
state.  Selection of this ground state then affects the electronic and
magnetic properties of FeSe, but only indirectly through the fact that
FeSe is in this particular state.  The direct effect of the SrTiO$_3$
on the electronic structure of an FeSe monolayer near the Fermi level
is negligible. For example, relaxing the structure of FeSe on
SrTiO$_3$ and then removing SrTiO$_3$ atoms from the calculation does
not affect the electronic structure near the Fermi level (see Fig.~1
in the supplement~\cite{supplement}).  Therefore to speed up the
calculation of the electron-phonon coupling, we perform calculations
on an isolated FeSe layer, without explicitly including SrTiO$_3$. To
avoid the shear instability in the FeSe monolayer from removing of
SrTiO$_3$, we reduce the value of parameter A in Eq.~\ref{eq:Af} from
$A_{\rm c}$ to $0.9A_{\rm c}$ and confirm that the electron-phonon
matrix elements are not affected by this simplification 
by carrying out full calculation (see Table.~1 in the
supplement\cite{supplement}).

\begin{figure}[!t]
\centering
\includegraphics{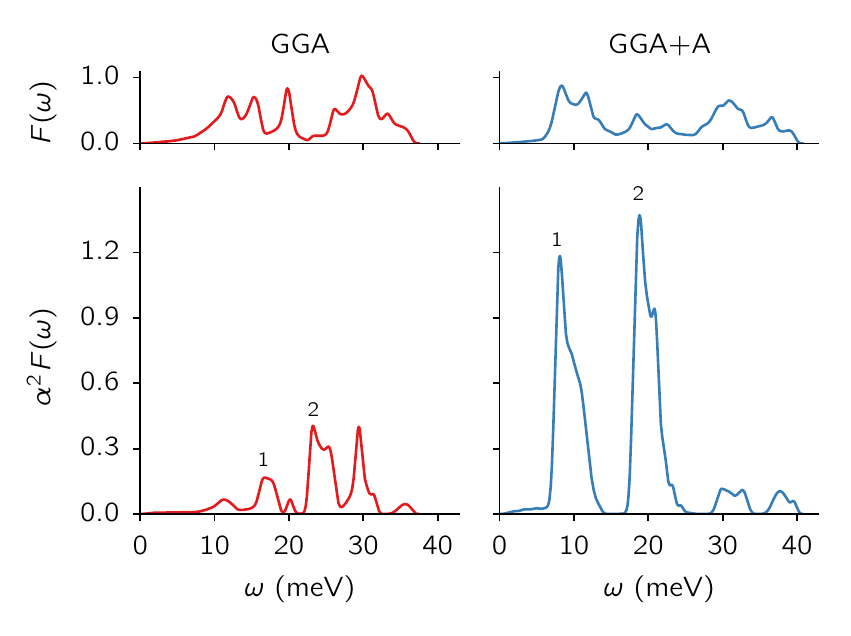}
\caption{\label{fig:a2F} Electron-phonon coupling $\alpha^2F(\omega)$
  and phonon density of states $F(\omega)$ (in meV$^{-1}$) in
  GGA and GGA+A (using $A=0.9 A_{\rm c}$).}
\end{figure}

We use state-of-the-art Wannier interpolation technique from
Ref.~\cite{giustino2007} and the Quantum-Espresso package described in
Ref.~\cite{QE-2009} to calculate the electron-phonon coupling in the
FeSe monolayer with a very fine grid in the Brillouin zone ($40 \times
40$). We obtained the superconducting transition temperature $T_{\rm
  c}$ by solving the Eliashberg
equation~\cite{migdal1958,eliashberg1960} as described in
Ref.~\cite{margine2013}. Figure~\ref{fig:a2F} shows the calculated
Eliashberg spectral function $\alpha^2F(\omega)$ of the FeSe
monolayer.  We focus our analysis on two groups of phonons for which
the electron-phonon coupling is the largest.  The first group of
phonons (labeled 1 in Fig.~\ref{fig:a2F}) corresponds to phonons with
frequency close to 10~meV, and the second group (labeled 2) to 
phonons with 20~meV
(in GGA those frequencies are 15 and 25~meV, respectively).

While phonons 1 contribute to about two-thirds of the total
electron-phonon coupling strength $\lambda$, they contribute to about
half of the integrated $\alpha^2F(\omega)$ spectral function (since
they have a lower frequency).

The atomic displacement character of the two groups of phonons is
different.  Phonons 1 correspond to a branch of phonons that involve
transverse, mostly in-plane displacements of atoms (these phonons
cause bulk FeSe to undergo a shear phase transition), while phonons 2
correspond to an out-of-plane transverse displacement of Fe atoms.
Furthermore, phonons 1 and 2 couple different parts of the electron
Fermi surface at M. Phonons 1 couple mostly at parts of the reciprocal
space where the Fermi surface (electron M pocket) crosses the
M--$\Gamma$ line and the least where it crosses the M--X line. The
opposite is true for phonons 2.  However, since both phonons
contribute about equally to $\alpha^2F$ the total electron-phonon
coupling (1 and 2 taken together) is nearly constant on the entire M
pocket Fermi surface.

\begin{table}[!t]
  \caption{\label{tab:lambdas} Electron-phonon coupling ($\lambda$),
    density of states (DOS), and average phonon frequencies in GGA and
    conservative estimates in GGA+A.}
\begin{ruledtabular}
\begin{tabular}{lccc}
& \mc{$\lambda$} 
& \mc{DOS}
& \mc{$\sqrt{\langle \omega^2 \rangle}$}
\\
&
& \mc{(eV$^{-1}$)}
& \mc{(K)}
\\
\hline
GGA          &0.3&1.0&252\\
GGA+A        &1.6&1.8&171\\
\end{tabular}
\end{ruledtabular}
\end{table}

Hence the importance of the SrTiO$_3$ substrate for increasing the
superconducting transition temperature within the phonon mechanism in
FeSe is two-fold. First, it prevents phonons 1 from becoming unstable
and induce a structural phase transition (as in bulk FeSe).  Second,
SrTiO$_3$ keeps FeSe in the checkerboard magnetic phase which allows
coupling of phonons from groups 1 and 2. In the non-magnetic case, the
coupling of these phonons is zero by symmetry.\cite{coh2015}
Calculations in Refs.~\cite{boeri2010,bazhirov2012} also found a
significantly smaller electron-phonon coupling in the non-magnetic
phase than in the magnetic phase. We also note that at this time,
there is no direct experimental measurement of magnetic order in FeSe
monolayer on SrTiO$_3$.  However, the measured ARPES band structure is
most closely resembled to that of the band structure of FeSe with an
antiferromagnetic checkerboard order, both in our GGA+A calculation
and in previous work.\cite{fawei2013,bazhirov2013} Nevertheless, it is
possible that the true ground state of FeSe monolayer consists of
fluctuating antiferromagnetic moments on iron atoms. Treatment of
electron-phonon coupling in such a state from first-principles goes
well beyond the scope of this work.

Comparing $\alpha^2F(\omega)$ in GGA and GGA+A (Fig.~\ref{fig:a2F}),
we find two reasons for an increased coupling in GGA+A. First,
preference for a shear distortion in GGA+A increases the
electron-phonon matrix elements of phonons 1 (see Fig.~3 in the
supplement~\cite{supplement}).  Second, the bottom of the electron M
pocket $E_{\rm M}$ is closer to the Fermi level in GGA+A than in the
GGA.  Therefore, owing to this band renormalization (narrowing of the
occupied bandwidth), the density of states at the Fermi level in GGA+A
is larger than in GGA (see Table~\ref{tab:lambdas} here and Fig.~3 in
the supplement~\cite{supplement}).  Since $\lambda$ is proportional to
the density of states, it is therefore increased in GGA+A.

However, as discussed earlier, we calculated the electron-phonon
coupling $\lambda$ within GGA+A with a reduced value of parameter $A$
from Eq.~\ref{eq:Af}.  Taking into account calculated density of
states (1.5~eV$^{-1}$) with $A=0.9A_{\rm c}$ and $A=A_{\rm c}$
(1.8~eV$^{-1}$) we conservatively estimate that the value of $\lambda$
at $A=A_{\rm c}$ is $\lambda=1.6$.  Next we use the Eliashberg theory
and obtain a conservative estimate of the superconducting transition
temperature $T_{\rm c}$ of 26~K (with $\mu^{*}=0.0$) and 21~K (with
$\mu^{*}=0.1$). This estimate is significantly closer to experiment
than a standard GGA result (0.1--1.5~K).

This range of estimated transition temperatures (21--26~K) is close to
the value found across the families of bulk iron-based
superconductors. Now we discuss possible reasons for an even larger
$T_{\rm c}$ in the case of an FeSe monolayer on SrTiO$_3$ (65--109~K).

When $\lambda$ is large, transition temperature is proportional
to~\cite{allen1975}
\begin{align}
T_{\rm c} \sim \overline{\omega} \lambda^{0.5}.
\label{eq:ad}
\end{align}
Here $\overline{\omega}$ is the averaged phonon frequency and
$\lambda$ is the Brillouin zone averaged electron-phonon coupling
strength. Therefore one possibility to get larger $T_{\rm c}$ is to
further increase $\lambda$. It is at least plausible that this could
happen for phonons 1, since their contribution to $\lambda$ is
increased when FeSe is approaching the shear-like structural phase
transition.

The second possibility is to increase the average frequency
$\overline{\omega}$ by pairing electrons with high frequency modes
(phonons or some other bosons) in addition to phonons 1 and 2. One
possibility are magnetic fluctuations~\cite{mazin2008}. The role of
magnetism for superconductivity in FeSe is additionally enriched by
the fact that, in the nonmagnetic phase, certain electron-phonon
interaction channels are forbidden by symmetry. In addition,
structural and magnetic order parameters are strongly coupled in FeSe.
For example, bulk orthorhombic FeSe prefers a non-magnetic state,
while a cubic FeSe monolayer on SrTiO$_3$ prefers an antiferromagnetic
state.

Another tempting possibility suggested in Ref.~\cite{jjlee2013} is to
pair FeSe electrons to a high-frequency (80~meV) phonon in the
SrTiO$_3$ substrate. This coupling was experimentally determined to be
large near the origin of the phonon Brillouin zone ($q\sim0$).
Adding experimentally estimated values of the electron-phonon coupling
from Ref.~\cite{jjlee2013} to our calculated $\alpha^2F(\omega)$
increases the estimated superconducting transition temperature to 47~K
(assuming $\mu^{*}=0.1$), even closer to the experimentally determined
value (65--109~K).

\begin{figure}[!t]
\centering
\includegraphics{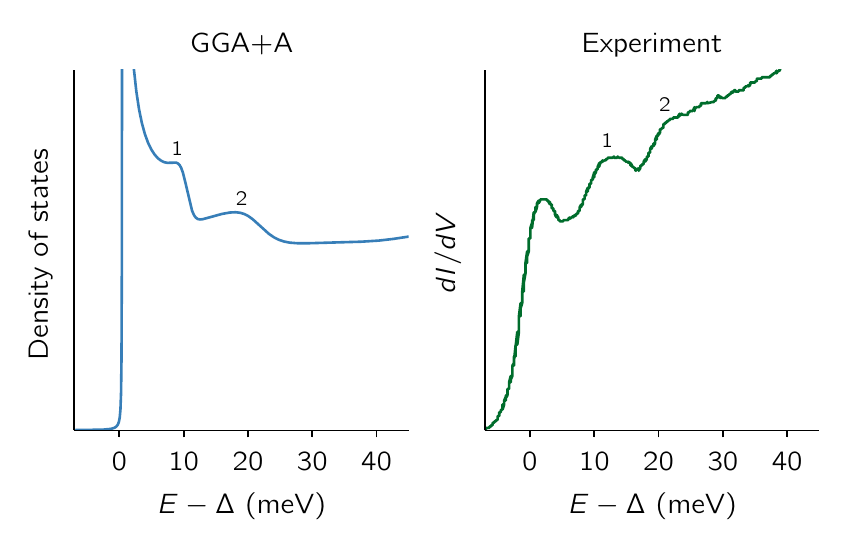}
\caption{\label{fig:stm} The density of states within the
    Eliashberg theory calculated using GGA+A and the STM measurement
    from Ref.~\cite{zhi2014}. The energy is measured relative to the
    superconducting gap $\Delta$.}
\end{figure}

In closing, we note that the experimentally inferred superconducting
$T_{\rm c}$ is nearly the same for an FeSe monolayer on TiO$_2$
terminated SrTiO$_3$~\cite{ge2014,wang2012},
BaTiO$_3$~\cite{peng2014}, as well as 2\% strained
SrTiO$_3$~\cite{pengPRL2014}. This observation is consistent with our
structural stabilization mechanism since in all three cases
interaction between Ti atoms in the TiO$_2$ layer and Se atoms in FeSe
is likely the same. However, when a FeSe monolayer is placed on a
substrate with a different bonding environment, such as SiC in
Ref.~\cite{song2011PRB,song2011,song2014} the superconducting T$_{\rm
  c}$ is only 2--9~K.  Another indication for the importance of
structural stabilization comes from Ref.~\cite{mcqueen2009}. This
study found that bulk FeSe doped with only 2\% of iron stays
tetragonal (non-sheared) even well below 90~K. This loss of preference
for shear is accompanied with loss of superconductivity ($T_{\rm
  c}<0.5$~K), again consistent with our finding that keeping FeSe
close to a shear (orthorhombic, nematic) structural phase transition
increases the electron-phonon coupling strength.  Another indication
of contribution from electron-phonon mechanism is described in
Ref.~\cite{khasanov2010} on iron isotope effect measurement.

Finally, our calculation is consistent with the inelastic scanning
tunneling microscope (STM) measurements from Ref.~\cite{zhi2014} in
two respects. First, the superconducting gap in the STM measurements
(as well as in ARPES in Refs.~\cite{liu2012,pengPRL2014}) is
node-less, just as is our calculated electron-phonon coupling being
nearly constant around the M pocket. Second, both our calculation and
the STM measurements find two peaks in the density of states above the
superconducting gap $\Delta$ (see Fig.~\ref{fig:stm}). One of these
peaks is at 10~meV and another at 20~meV above the gap.  As shown in
Ref.~\cite{mcmillan1965}, features in the tunneling spectrum above the
gap can be associated with $\alpha^2F$. Therefore, we tentatively
assign the two peaks found in the STM measurements to the strongly
electron-phonon coupled modes 1 and 2 discussed earlier in the text.

\begin{acknowledgments}
  This research was supported by the Theory Program at the Lawrence
  Berkeley National Lab through the Office of Basic Energy Sciences,
  U.S. Department of Energy under Contract No.  DE-AC02-05CH11231
  which provided for the electron-phonon calculation; and by the
  National Science Foundation under Grant No. DMR10-1006184 which
  provided for the structural and magnetic study. This research used
  resources of the National Energy Research Scientific Computing
  Center, which is supported by the Office of Science of the U.S.
  Department of Energy.
\end{acknowledgments}

\bibliography{pap}

\begin{thebibliography}{51}
\bibitem [{coh()}]{coh2015}%
  \BibitemOpen
  \href@noop {} {}\bibinfo {note} {Coh S, Cohen M L and Louie S G, to be
  published.}\BibitemShut {Stop}%
\end{thebibliography}



\section*{Supplementary material}

# Supplemental material for "Large electron-phonon interactions from FeSe phonons in a monolayer"

Sinisa Coh, Marvin L. Cohen, and Steven G. Louie

*Department of Physics, University of California at Berkeley and Materials Sciences Division,*
*Lawrence Berkeley National Laboratory, Berkeley, California 94720, USA*
(Dated: July 10, 2015)

*Technical details.* For the semi-core iron pseudopotential used we obtain $A_c f(r) = (5.7 \text{ Ry}) e^{-r^2/(0.85 \text{ bohr})^2}$. Near the center of the atom, the depth of this potential is only 6% of the local part of the pseudopotential. Throughout this work, we use GGA-PBE functionals with norm-conserving pseudopotentials that include semi-core electrons on Fe and Ti. We use a 180 Ry energy cutoff, 8x8x1 sampling of the electron Brillouin zone, and 4x4x1 sampling of the phonon Brillouin zone which we Wannier interpolate on a 40x40x1 grid. All conventions in the paper and the supplement are for the primitive unit cell with two Fe atoms per cell. Electron doping in our calculations equals 0.09 electrons per one Fe atom (as found in ARPES experiment).

*Additional figures and one table can be found on the following pages.*

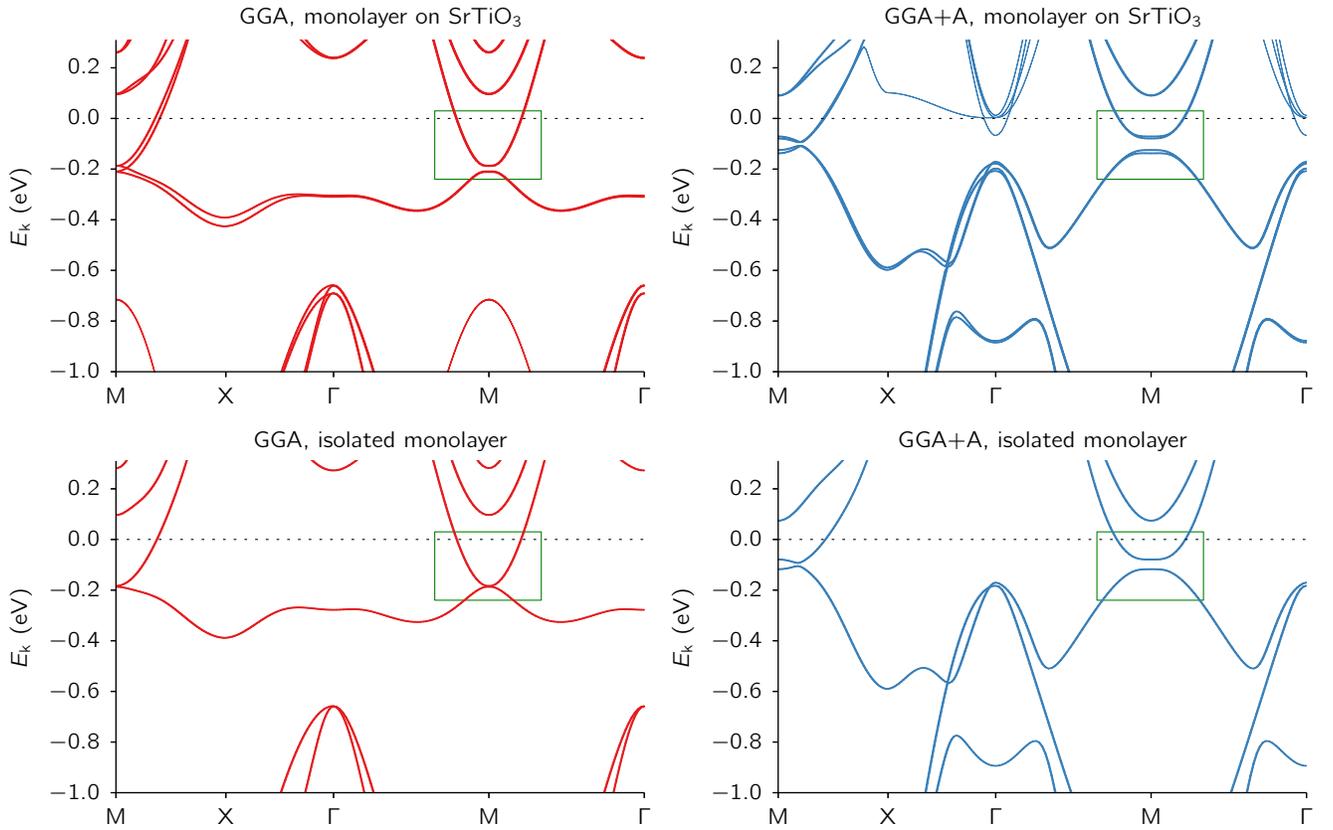

FIG. 1. Electronic band structure of FeSe monolayer on a path in momentum space calculated in GGA (red) and GGA+A (blue). Conventions are for the primitive unit cell with two Fe atoms per cell. Top two panels show the band structure of FeSe monolayer on a $SrTiO_3$ slab after a full structural relaxation. Bottom panels show the band structure when $SrTiO_3$ atoms are removed from the calculation while the FeSe monolayer atoms are kept at the same positions as in the full calculation. Thin lines in the topmost panels correspond to the electronic bands localized on the $SrTiO_3$ slab. Thick lines are states localized on FeSe. Green squares indicate region of the band structure plot that is shown in Fig. 1 in the main text. Computational unit cell used in our calculation consists of a $SrTiO_3$ slab covered with FeSe monolayer on each end, so that both ends of $SrTiO_3$ are passivated in the same manner. Hybridization between these two monolayers of FeSe causes small splitting seen on the M-X line at 0.2 eV below the Fermi level in the top-left panel, and at the bottom of the M pocket in the top right panel. This hybridization splitting disappears in the lower panels because in that case computational cell contains only one FeSe monolayer.

7

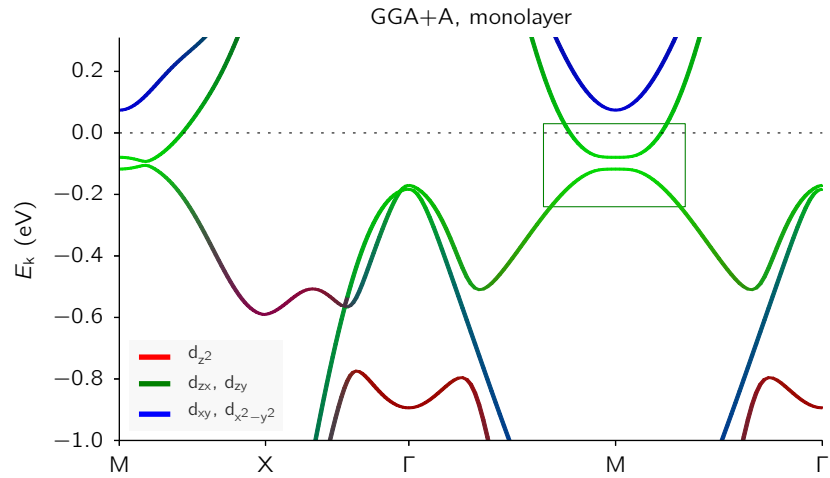

FIG. 2. The d-character of the electron wavefunction in the FeSe monolayer. Line segments colored red corresponds to the $d_{z^2}$ states, green to the $d_{zx}$ and $d_{zy}$, and blue to $d_{xy}$ and $d_{x^2-y^2}$ states. Everything in the figure except for the coloring scheme is the same as in the lower right panel of Fig. 1 in the supplement.

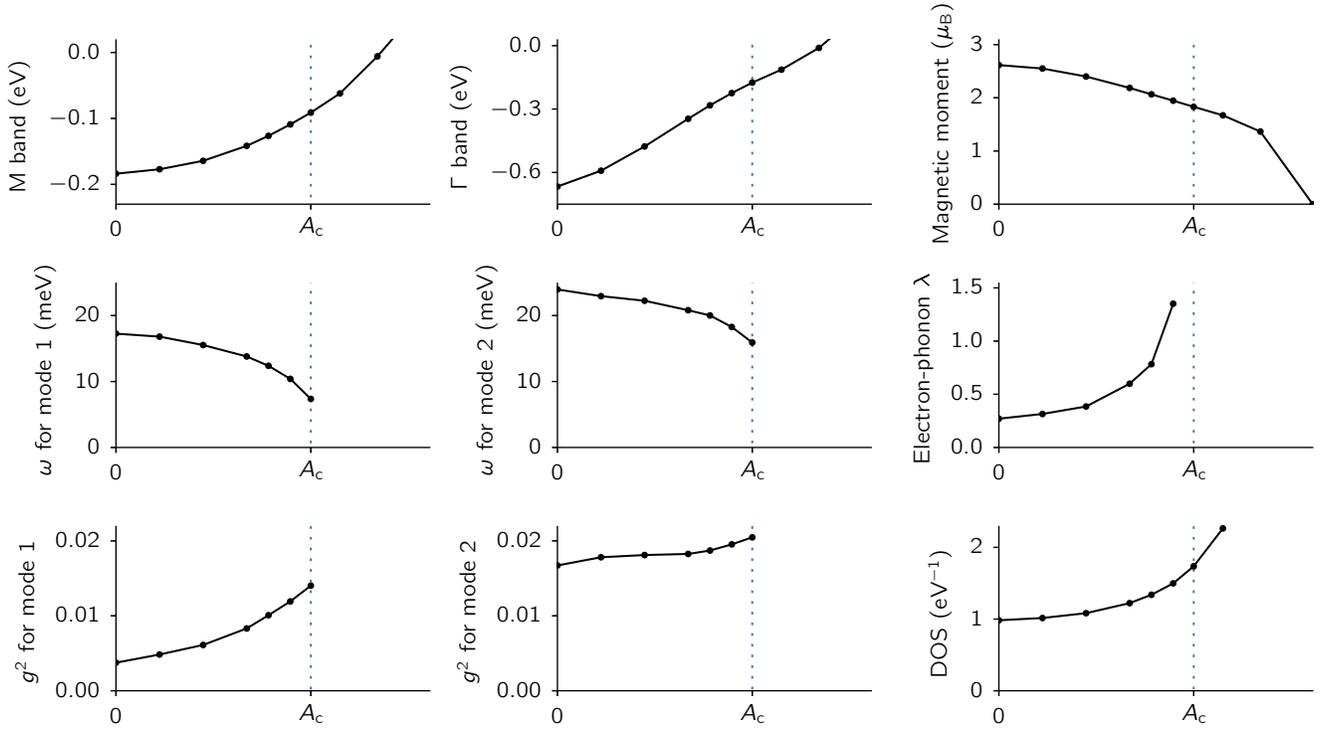

FIG. 3. Dependence of nine properties of an isolated FeSe monolayer on the value of parameter A in the GGA+A approach: bottom of the M pocket (occupied bandwidth, $E_M$), top of the $\Gamma$ pocket ($E_\Gamma$), magnetic moment on the iron atom, frequency of phonons ($\omega$) with dominant electron-phonon coupling (modes 1 and 2 discussed in the main text), total electron-phonon coupling strength $\lambda$, electron phonon matrix element squared ($g^2$) for dominant modes 1 and 2, and density of states at the Fermi level. Dashed blue line shows $A = A_c$ case discussed in the main text. Regular GGA result (without A) corresponds to $A = 0$. Here the SrTiO$_3$ substrate was not explicitly included in the calculation. Instead, the FeSe monolayer was structurally relaxed with the in-plane lattice constant equal to that of SrTiO$_3$. Phonon eigendisplacements of modes 1 and 2 are taken from $A = 0.9 A_c$ calculation and then used for all values of $A$ for consistency.

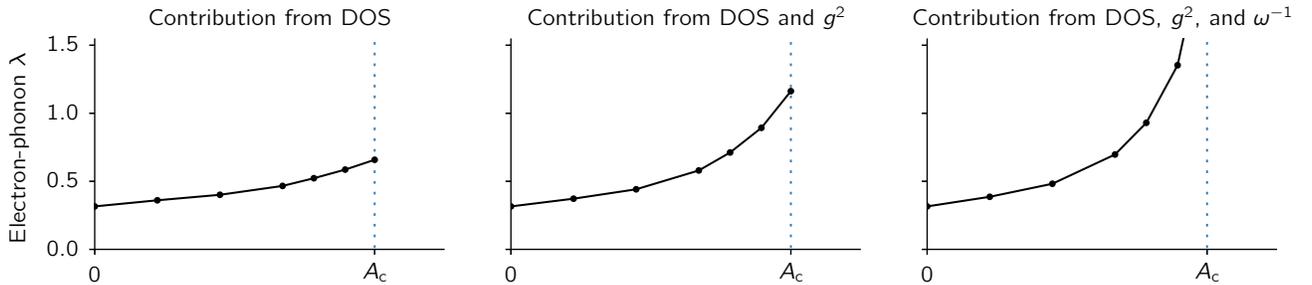

FIG. 4. Contribution to lambda in the case of an isolated FeSe monolayer. We show contributions of density of state (DOS) increase, electron-phonon matrix element squared ($g^2$) increase, and phonon frequency softening ($\omega^{-1}$) to the increase in $\lambda$ as a function of parameter $A$. Critical behavior of $\lambda$ at $A_c$ originates from phonon softening at $A = A_c$ since these calculations do not include SrTiO$_3$ substrate. As we show in Table. I and Fig. 5, critical behavior is removed once substrate is included in the calculation. Here we used calculated contribution of phonons 1 and 2 to $\lambda$ at $0.9 A_c$ and isolated dependence of $\lambda$ on DOS, $g^2$, and $\omega^{-1}$ using the following simplified expression for $\lambda \sim \mathrm{DOS} \times g^2 \times \omega^{-1}$.

TABLE I. Electron-phonon matrix elements ($g^2$ in atomic units) and phonon frequencies (in meV) of the dominating phonon modes (modes 1 and 2, see main text) at $A = 0.9A_c$ calculation without a substrate, and $A = A_c$ with a substrate. For computational convenience, we first calculate phonon eigendisplacements with $A = 0.9A_c$ and no substrate, and then use those same eigenvectors in the $A = A_c$ calculation explicitly including the SrTiO$_3$ substrate.

|  | Dominant phonon 1 | | Dominant phonon 2 | |
| --- | --- | --- | --- | --- |
|  | $g^2$ (au) | $\omega$ (meV) | $g^2$ (au) | $\omega$ (meV) |
| Isolated FeSe with $A = 0.9A_c$ | 0.012 | 10 | 0.020 | 18 |
| FeSe on SrTiO$_3$ substrate with $A = A_c$ | 0.010 | 7 | 0.019 | 17 |

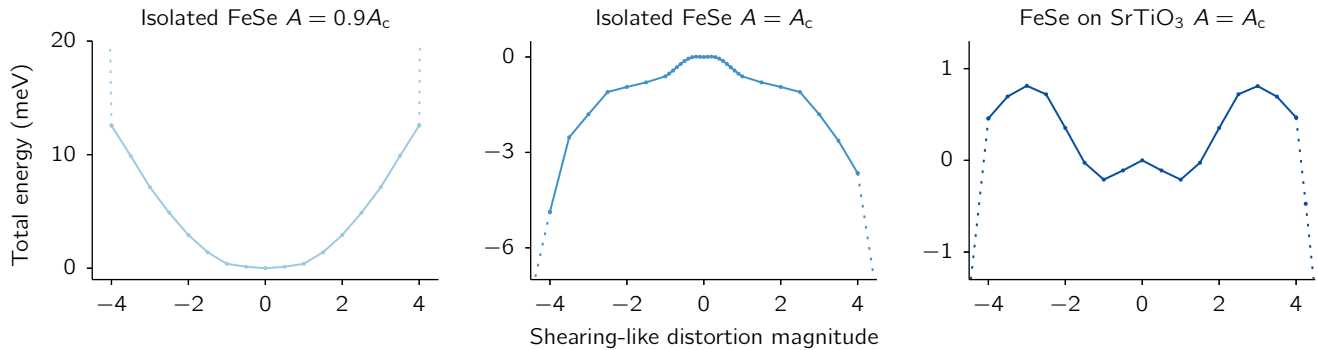

FIG. 5. Total energy per unit cell (Fe$_2$Se$_2$ formula unit) as a function of shearing-like distortion for isolated FeSe with $A = 0.9A_c$ and $A = A_c$, as well as FeSe on SrTiO$_3$ with $A = A_c$. Presence of SrTiO$_3$ stabilizes the shearing-like distortion present in the isolated FeSe monolayer at $A_c$. Only a small amount of shearing distortion is present in the fully relaxed ground state of FeSe on SrTiO$_3$ (corresponding to $\pm 1$ on the horizontal scale). This distortion is responsible for a gap ($\delta E_M$) at the $M$ point shown in Fig. 1 of the main text. If shearing distortion is artificially increased beyond $\pm 4$ (dashed lines), the gap $\delta E_M$ increases and the system undergoes a metal-insulator transition in all three cases. Therefore, the energy landscape for a shearing distortion in FeSe monolayer is driven by a gap opening of a small-area electron pocket at the M point.

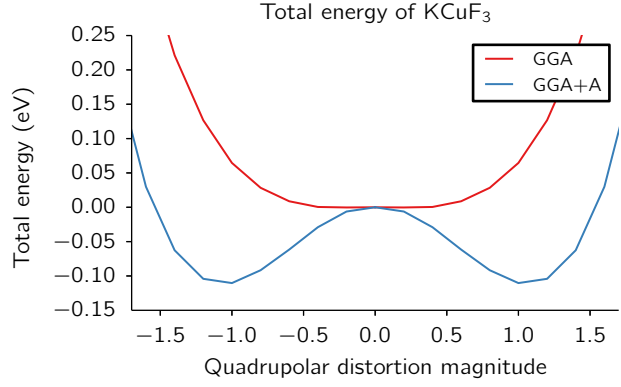

FIG. 6. Total energy of paramagnetic KCuF$_3$ (per one formula unit) as a function of fluorine quadrupolar distortion within GGA (red) and GGA+A (blue). Ground state within GGA shows no preference for this distortion unlike GGA+A (horizontal scale is chosen so that ±1 corresponds to the experimental value of the distortion magnitude). Similar preference for fluorine quadrupolar distortion was found by including Hubbard +U and +J term in Ref. [1].

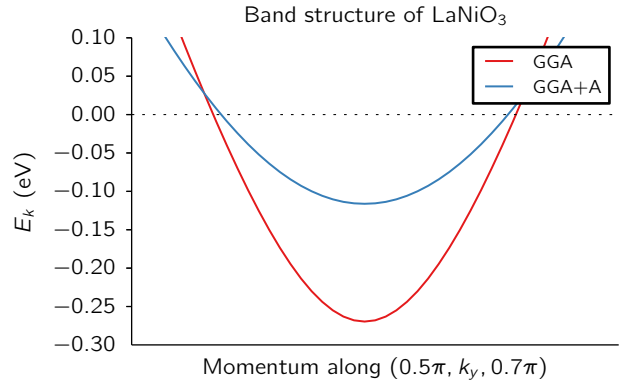

FIG. 7. Band structure of LaNiO$_3$ along the $(0.5\pi, k_y, 0.7\pi)$ line shows a large mass renormalization once +A term is included (blue versus red curve). This is consistent with the experimental finding from Ref. [2] (ARPES data in Ref. [2] is shown along the same path in the momentum space).

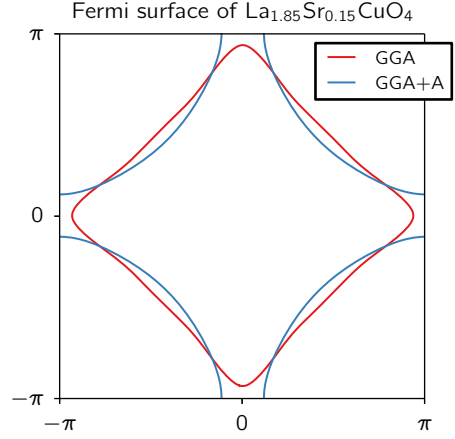

FIG. 8. Fermi surface of optimally doped $La_{1.85}Sr_{0.15}CuO_4$ shows different topology in GGA (red) and GGA+A (blue), with the latter consistent with experimental findings in Ref. [3].

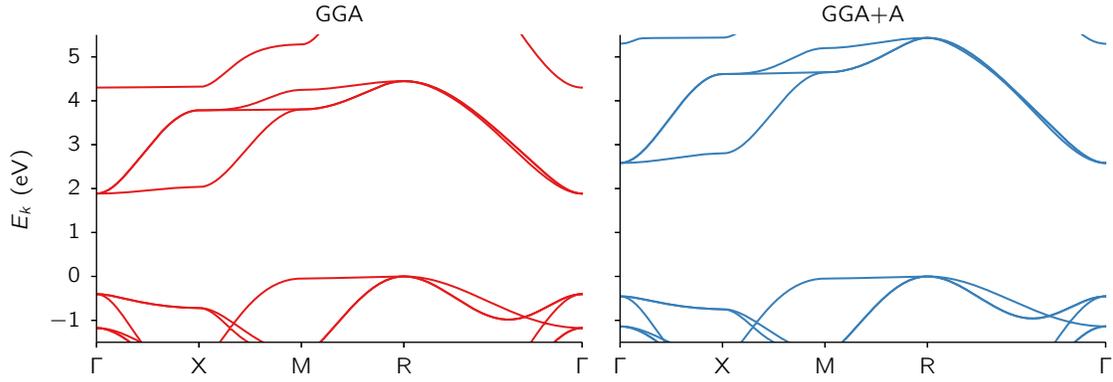

FIG. 9. Band structure of $SrTiO_3$ within GGA (red) and GGA+A (blue). Direct and indirect band gaps in GGA (2.3 and 1.9 eV) are increased once the +A term is included (3.0 and 2.6 eV) in better agreement with the experimental data from Ref. [4].



\end{document}